\pdfoutput=1
\RequirePackage{ifpdf}
\ifpdf 
\documentclass[pdftex]{sigma}
\else
\documentclass{sigma}
\fi

\numberwithin{equation}{section}
\DeclareMathOperator{\rank}{\mathrm{rank}}
\DeclareMathOperator{\ad}{ad}
\DeclareMathOperator{\Ad}{Ad}
\newcommand{\DD}[2]{\frac{\partial #1}{\partial #2}}
\newcommand{\D}[1]{\frac{\partial}{\partial #1}}

\begin{document} 	

\newcommand{\arXivNumber}{1312.0362}


\renewcommand{\PaperNumber}{066}

\FirstPageHeading

\ShortArticleName{Computation of Composition Functions and Invariant Vector Fields}

\ArticleName{Computation of Composition Functions\\
and Invariant Vector Fields in Terms of Structure\\
Constants of Associated Lie Algebras}

\Author{Alexey A.~MAGAZEV, Vitaly V.~MIKHEYEV and Igor V.~SHIROKOV}
\AuthorNameForHeading{A.A.~Magazev, V.V.~Mikheyev and I.V.~Shirokov}
\Address{Omsk State Technical University, 11 Mira Ave., Omsk, 644050, Russia}
\Email{\href{mailto:magazev@gmail.com}{magazev@gmail.com}, \href{mailto:vvm125@mail.ru}{vvm125@mail.ru}, \href{mailto:iv_shirokov@mail.ru}{iv\_shirokov@mail.ru}}

\ArticleDates{Received December 05, 2013, in f\/inal form July 25, 2015; Published online August 06, 2015}

\Abstract{Methods of construction of the composition function, left- and right-invariant vector f\/ields and dif\/ferential
1-forms of a Lie group from the structure constants of the associated Lie algebra are proposed.
It is shown that in the second canonical coordinates these problems are reduced to the matrix inversions and matrix
exponentiations, and the composition function can be represented in quadratures.
Moreover, it is proven that the transition function from the f\/irst canonical coordinates to the second canonical coordinates
can be found by quadratures.}


\Classification{22E05; 22E60; 22E70}

\section{Introduction}

Researchers in the f\/ield of theoretical and mathematical physics who use methods of  Lie theory face the problem on
realizations of a~f\/inite-dimensional Lie algebra by means of vector f\/ields on a~certain domain of a f\/inite-dimensional real space.
This problem is vitally important for group classif\/ication of partial dif\/ferential equations~\cite{BasLahZhd01,
HerOlv96}, for the classif\/ication of pseudo-Rieman\-nian metrics on manifolds with groups of motions~\cite{Pet66}, as well as
for the construction of relativistic wave equations in external f\/ields with a~given symmetry
group~\cite{KurShi08}.
The more general and interesting problem on realizations of Lie algebras by nonhomogeneous f\/irst-order dif\/ferential
operators should also be mentioned in this context.
It naturally emerges in the theory of projective representations of Lie groups~\cite{BarShi03, Mil68}.
This problem is of great importance in applications, for instance in quantum theory of scattering~\cite{AlhEngWu84} and in
integration of dif\/ferential and integro-dif\/ferential equations~\cite{GonShi09, ShaShi95}.

The problem on realizations of a Lie algebra by vector f\/ields has long history and goes back to works of S.~Lie but
modern mathematicians still demonstrate their interest to this f\/ield and their approaches are directly depend on applications.
As a~result, now there are a~suf\/f\/iciently large number of works in this f\/ield.
We point out some of the most important results.

\looseness=-1
Undoubtedly, S.~Lie stated the principal ideas in this f\/ield, and the f\/irst important results also belong to him.
For example, he listed all possible realizations of f\/inite-dimensional Lie algebras on the real and
complex lines. Later he presented the similar result for the complex plane~\cite{Lie80}.
The results of S.~Lie were completed by the classif\/ication of vector f\/ields on the two-dimensional real
plane~\cite{GonKamOlv92}.
Further ef\/forts of mathematicians were mainly concentrated on the classif\/ication of realizations of
low-dimensional Lie algebras.
Here we would like to point out the important paper~\cite{PopBoyNesLut03}, where the special technique of so called
\textit{mega\-ideals} was used to list all inequivalent realizations of Lie algebras up to dimension four by vector
f\/ields on an arbitrary real (resp.\ complex) f\/inite-dimensional space.
(This paper also contains a~quite complete list of references on the discussed problem.)
At the same time, a~number of researchers classif\/ied inequivalent realizations of Lie algebras that are important for
theoretical physics.
Such classif\/ications were done for the Lie algebras of the Euclidean group~${\rm E}(3)$ and the Poincar\'e group~${\rm P}(1,3)$
(see, e.g.,~\cite{FusZhd97}).
Some important results were also obtained for some inf\/inite series of Lie groups and algebras.
For instance, the constructive algorithm of embedding of an arbitrary $\mathbb{Z}$-graded Lie algebra into a~Lie algebra of
polynomial vector f\/ields over a f\/ield of arbitrary characteristic was described in~\cite{Shc06}.
We also have to mention the review paper~\cite{Dra12} in which the author consider the problem on realizations of
transitive Lie algebras by formal vector f\/ields.

In the present paper we introduce a~method to construct an explicit realization of a~f\/inite-dimensional Lie algebra by left- and
right-invariant vector f\/ields on the associated local Lie group using only the structure constants of the algebra.
It is shown that in the second canonical coordinates the problem can be solved just by tools of linear algebra
and it is reduced to the computation of matrix inversions and matrix exponentiations.
The introduced method allows one to construct only regular realizations of Lie algebras but its possible applications are
wider.
Indeed, if we can construct the realization of the Lie algebra by left-invariant vector f\/ields in canonical coordinates,
then we can list other inequivalent realizations of this algebra by vector f\/ields depending on smaller number of
independent variables.
This can be done by the classif\/ication of inequivalent subalgebras of the initial Lie algebra and by the projection of the
left-invariant vector f\/ields on the corresponding spaces of right cosets.

Even more complicated problem is solved below.
This is the construction of the composition function of a local Lie group whose Lie algebra is known.
We emphasize this problem since knowing the composition function gives the complete description of the
group structure.
Mo\-dern approaches to the computation of composition functions is reviewed in the next section, but the
main result of this paper can be announced here:
The composition function in the second canonical coordinates can be found by quadratures.
In Section~\ref{Section5} it is shown that the transition from the second canonical coordinates to the f\/irst canonical coordinates
can be found by quadratures too.
Therefore, if one knows the composition function in any system of canonical coordinates, then the transition to another
system of canonical coordinates can be done using special techniques described in this paper.

\section{Preliminary information on theory of Lie groups and algebras}\label{Section2}

To make the presentation self-contained and to f\/ix the notations we present basic facts of the Lie theory.

Let $G_e$ be an open neighborhood of the identity element~$e$ of an~$n$-dimensional simply connected real Lie group~$G$
that is dif\/feomorphic to an open subset~$U$ of the Euclidean space~$\mathbb{R}^n$ and let
$\psi$~be a mapping realizing this dif\/feomorphism, $\psi\colon G_e\to U$.
Any group element from the domain~$G_e$ is uniquely def\/ined by its coordinates.
Explicitly this can be expressed as $g_x=\psi^{-1}(x)\in G_e$, where $x=(x^1,\dots,x^n)\in U$.
Therefore, the multiplication rule is represented~as\footnote{It is clear that for all objects to be well def\/ined in~\eqref{e1-1-1},
we should consider only pairs of~$x$ and~$y$ with $g_xg_y\in G_e$.
In what follows we omit similar conditions for coordinates.}
\begin{gather}
\label{e1-1-1}
g_x g_y = g_z,
\qquad
z^i = \Phi^i(x,y),
\qquad
g_x, g_y, g_z \in G_e.
\end{gather}
The $n$-dimensional vector function $\Phi(x,y) = (\Phi^1(x,y), \dots, \Phi^n(x,y))$ is called a~\textit{composition function} of the group~$G$.
Since the group multiplication is associative, the function~$\Phi(x,y)$ satisf\/ies the identity
\begin{gather*}
\Phi(x,\Phi(y,z))=\Phi(\Phi(x,y),z).
\end{gather*}
Without loss of generality one can assume that the zero value of the coordinate tuple corresponds to the identity element~$e$ of the group, $\psi(e)=0$.
Then the composition function satisf\/ies the initial conditions
\begin{gather}
\label{e1-1-3}
\Phi(0,y) = y,
\qquad
\Phi(x,0) = x.
\end{gather}
Denote by $\kappa(x)$ the coordinates of the inverse of $g_x$, so $g_x^{-1} = g_{\kappa(x)}$.
Then the following equalities are obvious:
\begin{gather*}
\Phi(\kappa(x), x) = \Phi(x,\kappa(x)) = 0,
\qquad
\left.\DD{\kappa^i(x)}{x^j}\right|_{x = 0} = -\delta^i_j.
\end{gather*}

The tangent vectors $\partial_{x^i}$ at a point~$x$ constitute a~basis of the tangent space $T_x\mathbb{R}^n$.
The corresponding tangent vectors $(\psi^{-1})_* \partial_{x^i}\equiv \partial_{x^i}g_x\in T_{g_x}G$ form a~basis of the tangent
space of the group~$G$ at the point $g_x$.
If we assume $x=0$ then the tangent vectors $\partial_{x^i}g_x|_{x=0}^{}\equiv e_i$ form a~basis of the Lie algebra
$\mathfrak{g}$ of the Lie group~$G$ with commutation relations
\begin{gather}
\label{e1-1-4}
[e_i,e_j]=C^k_{ij} e_k.
\end{gather}
The numbers $C_{ij}^k$ are the \textit{structure constants} of the Lie algebra~$\mathfrak{g}$ in the basis chosen.
Hereafter, we follow the Einstein summation convention assuming summation over the repeated indices unless otherwise stated.

The group~$G$ acts on itself by the right $R_{g_y}$ and the left $L_{g_y}$ translations,
\begin{gather*}
R_{g_y}g_x=g_xg_y,
\qquad
L_{g_y}g_x=g_yg_x.
\end{gather*}
These actions generate the right $T^R$ and the left $T^L$ regular representations of the group $G$,
\begin{gather*}
T^R(g_y) f(g_x) = f(g_x g_y),\quad
T^L(g_y) f(g_x) = f\big(g_y^{-1}  g_x\big),\quad
f \in C^\infty(G),
\end{gather*}
whose generators are left- and right-invariant vector f\/ields, respectively,
\begin{gather}
\xi_i(g_x)=(L_{g_x})_*e_i\in T_{g_x}G,
\qquad
\psi_*\xi_i(g_x)\equiv\xi_i(x)=\xi^j_i(x)\D{x^j},
\nonumber
\\
\eta_i(g_x)=-(R_{g_x})_*e_i\in T_{g_x}G,
\qquad
\psi_*\eta_i(g_x)\equiv\eta_i(x)=\eta^j_i(x)\D{x^j},
\nonumber
\\
\label{e1-1-5}
\xi^j_i(x)=\left.\DD{\Phi^j(x,y)}{y^i}\right|_{y=0},
\qquad
\eta^j_i(x)=\left.\DD{\Phi^j(\kappa(y),x)}{y^i}\right|_{y=0}=-\left.\DD{\Phi^j(y,x)}{y^i}\right|_{y=0}.
\end{gather}
By $\|\xi(x)\|$ and $\|\eta(x)\|$ we denote the matrices formed by the components $\xi_i^j(x)$ and $\eta_i^j(x)$, respectively.
The left- and right-invariant basis vector f\/ields $\xi_i$ and $\eta_j$ satisfy the commutation relations~\eqref{e1-1-4} and
commute with each other, $[\xi_i,\eta_j]=0$.

Denote by $\omega^i$ and $\sigma^i$ the dif\/ferential 1-forms on~$G$ that are dual to the vector f\/ields $\xi_i(g_x)$ and
$\eta_i(g_x)$, respectively, $\langle\omega^i,\xi_j\rangle=\langle\sigma^i,\eta_j\rangle=\delta^i_j$.
The basis 1-forms can be written in coordinates~$x^i$~as
\begin{gather}
\label{e1-1-5b}
\omega^i=\omega^i_j(x)dx^j,
\qquad
\omega^i_j(x)=\big\|\xi^{-1}(x)\big\|^i_j,
\qquad
\sigma^i=\sigma^i_j(x)dx^j,
\qquad
\sigma^i_j(x)=\big\|\eta^{-1}(x)\big\|^i_j.
\end{gather}
The commutation relations~\eqref{e1-1-4} can be rewritten in the form of equations for the invariant 1-forms
\begin{gather}
\label{e1-1-5a}
d\omega^i=-\frac12 C^i_{jk} \omega^j\wedge \omega^k,
\qquad
d\sigma^i=-\frac12 C^i_{jk} \sigma^j\wedge \sigma^k.
\end{gather}
These equations are known as \textit{Maurer--Cartan equations}.

An element~$g$ of the group~$G$ generates the inner automorphism $\Ad_g$ of the algebra~$\mathfrak{g}$,
\begin{gather}
\label{e1-1-6}
\Ad_{g} e_i\equiv (L_{g})_*(R_{g^{-1}})_* e_i=\|\Ad_{g}\|_i^j e_j,
\qquad
\D{x^k} \|\Ad_{g_x}\|_i^j\bigg|_{x = 0} = C_{ki}^j = \|\ad e_k\|^j_i.
\end{gather}
The matrix $\Ad_{g_x}$ can be expressed in terms of the components of basis invariant vector f\/ields and  dual 1-forms
\begin{gather}
\label{e1-1-6b}
\|\Ad_{g_x}\|^i_j=-\sigma^i_k(x)\xi^k_j(x),
\end{gather}
so that $\Ad_{g_x}=-\|\sigma(x)\|\cdot\|\xi(x)\|$.

The composition function satisfying the initial conditions~\eqref{e1-1-3} can be uniquely determined from the condition
of left- or right-invariance of the vector f\/ields $\xi_j$ and $\eta_j$, respectively,
\begin{gather}
\label{e1-1-7}
\DD{\Phi^k(x,y)}{y^i}=\xi^k_j(\Phi(x,y))\omega^j_i(y),
\\
\label{e1-1-8}
\DD{\Phi^k(x,y)}{x^i}=\eta^k_j(\Phi(x,y))\sigma^j_i(x).
\end{gather}

Note that the above relations hold for any coordinate system on the Lie group~$G$.
Consider now special kinds of local coordinates.

Let the Lie algebra~$\mathfrak{g}$ (as a~vector space) be decomposed into a~direct sum of subspaces,
\begin{gather}
\label{e1-2-1}
\mathfrak{g} = \bigoplus \limits_{k = 1}^m \mathfrak{g}_k = \mathfrak{g}_1 \oplus \mathfrak{g}_2 \oplus \dots \oplus
\mathfrak{g}_m.
\end{gather}
We def\/ine a~mapping $\phi\colon \mathfrak{g} \rightarrow G$ by $\phi(X) = \prod
\limits_{k = 1}^m \exp(X_k)$ for any $X\in \mathfrak{g}$, where \mbox{$\exp\colon \mathfrak{g} \rightarrow G$} is the exponential map, $X_k$ is the component
of~$X$ corresponding to $\mathfrak{g}_k$ in the decomposition~\eqref{e1-2-1}.
There exists a~neighborhood~$U$ of $0 \in \mathfrak{g}$ such that~$\phi$ is a~dif\/feomorphism on a~neighbor\-hood~$G_e$ of
the identity element $e \in G$.
Therefore, the pair $(G_e, \phi^{-1})$ is a~map on~$G$, which is called \textit{canonical} and the respective local
coordinates $x = (x^1, \dots, x^n) \in U$ are called \textit{canonical coordinates}.

Consider two types of canonical coordinates that are frequently used~\cite{Che46, Coh65}.
Let the decomposition~\eqref{e1-2-1} be trivial, i.e., $m=1$ and $\mathfrak{g}_1 = \mathfrak{g}$. Then
\begin{gather*}
g_x = \exp \left(\sum \limits_{i = 1}^n x^i e_i \right) = \exp\big(x^1 e_1 + \dots + x^n e_n\big),
\end{gather*}
where $e_1, \dots, e_n$ constitute a~basis of the Lie algebra~$\mathfrak{g}$.
In this case, the canonical coordinates are called \textit{first canonical coordinates}.
If $m=n$ and thus the subspaces $\mathfrak{g}_k$ are necessarily one-dimensional,
then one has \textit{second canonical coordinates},
\begin{gather*}
g_x = \prod \limits_{i = 1}^n \exp\big(x^i e_i\big) = \exp\big(x^1 e_1\big) \cdots \exp\big(x^n e_n\big)
\end{gather*}
(the Einstein summation convention is not implied).
The choice of canonical coordinates depends on the problem to be solved.

One of classical problems of the theory of Lie groups is the construction of group multiplication law of a Lie group from the
structure constants of the associated Lie algebra.
The traditional way consists of two steps in accordance with the well-known Lie theorems.
First, the Maurer--Cartan equations~\eqref{e1-1-5a} are to be solved (to be specif\/ic we consider the f\/irst
system), where
the compo\-nents~$\omega_i^j(x)$ of the left-invariant 1-forms are assumed the unknowns.
Note that the obvious initial condition $\omega_i^ j(0) = \delta_i^j$ does not guarantee the solution uniqueness.
Therefore, at the f\/irst stage one usually chooses and f\/ixes a~certain system of canonical coordinates on the Lie group.
For instance, the equality $\omega^i_j(x) x^j = x^i$ holds true in the f\/irst canonical coordinates.
Using this equality the solution of~\eqref{e1-1-5a} can be represented in the explicit form~\cite{Pon73}
\begin{gather}
\label{e1-2-2a}
\Vert \omega(x) \Vert = \Omega(\ad_x),
\qquad
\Omega(s) = \frac{1 - e^{- s}}{s}.
\end{gather}
The second step of the computation of the composition function $\Phi(x,y)$ is the integration of the
equation~\eqref{e1-1-7} with the initial condition~\eqref{e1-1-3}.

Another possible way to construct the composition function for a~given Lie group from the structure constants of its Lie
algebra, requires the explicit computation of the element $Z = \ln\big(e^X e^Y\big)$ in the form
\begin{gather}
\label{e1-2-2}
Z = Y + \int_0^t \theta\big(e^{t \ad_X} e^{\ad_Y}\big) X dt,
\end{gather}
where $\theta(s) = \ln s /(s - 1)$, and the elements~$X$ and~$Y$ belong to a~suf\/f\/iciently small neighborhood of the zero element of the Lie
algebra~\cite{Mos78}.
Indeed, if $X = x^i e_i$ and $Y = y^i e_i$ are the decompositions of vectors~$X$ and~$Y$ in the f\/ixed basis of the Lie
algebra~$\mathfrak{g}$, respectively, then the components of the vector~$Z$ computed by formula~\eqref{e1-2-2} are the components of the
composition function $\Phi(x,y)$ in the f\/irst canonical coordinates.
One of the consequences of~\eqref{e1-2-2} is the Baker--Campbell--Hausdorf\/f series which can be obtained by means of the
decomposition of the function $\theta(s)$ in power series at the point $s = 1$.

Although the presented methods make it possible to f\/ind the group multiplication from the commutation relations of the
associated Lie algebra, they are not convenient for applications.
The f\/irst method based on the Lie theorems requires the integration of systems of nonlinear partial dif\/ferential
equations, which appears to be a quite dif\/f\/icult task even for low-dimensional Lie groups.
The use of~\eqref{e1-2-2} involves the complicated calculation of functions depending on matrices as
variables (details are discussed below).

Consider a~more promising way to construct the composition function.
Denote by $\mathrm{Mat}_m(\mathbb{R})$ the set of all square~$m\times m$ matrices over the f\/ield $\mathbb{R}$.
Let $\tau\colon \mathfrak{g}\to \mathrm{Mat}_m(\mathbb{R})$ be a~faithful f\/inite-dimensional representation of the Lie  
algebra~$\mathfrak{g}$.
Denote the neighborhood of the identity element in the group~$G$ as $U \subset G$.
Then~$V$ will stand for the neighborhood of the zero element of the Lie algebra~$\mathfrak{g}$, which is mapped onto~$U$
under the action of exponential mapping.
Then a~mapping~$T$ def\/ined as
\begin{gather*}
T(\exp X)=\exp(\tau(X)),
\qquad
X \in V,
\end{gather*}
gives locally homomorphic mapping of~$G$ into $\mathrm{GL}_m(\mathbb{R})$~\cite{Che46,HauSch68}.
It means that there exists such a~neighborhood of the identity element $G_e \subset U$ that $T(g_1 g_2) = T(g_1) T(g_2)$
for any $g_1, g_2 \in G_e$.
Replacing the group elements in~\eqref{e1-1-1} by their representations, $g\to T(g)$, results in the matrix
equality that can be used for the identif\/ication of all components of the composition function $z^i=\Phi^i(x,y)$; this
can be done as far as the representation~$\tau$ is faithful.
For example, in the f\/irst and the second canonical coordinates we get the following matrix equalities (there is no summation
over the repeated indices in the second formula)
\begin{gather*}
\exp\left(\sum \limits_{i = 1}^n x^i \tau(e_i)\right) \exp\left(\sum \limits_{j=1}^n y^j \tau(e_j)\right)=\exp\left(\sum
\limits_{k=1}^n z^k \tau(e_k)\right),
\\
\prod \limits_{i=1}^n\exp\big(x^i \tau(e_i)\big)\prod \limits_{j=1}^n\exp\big(y^j \tau(e_j)\big)=\prod
\limits_{k=1}^n \exp\big(z^k \tau(e_k)\big).
\end{gather*}

The main disadvantage of the present approach is the absence of a~simple procedure that allows to construct a~faithful
representation of an arbitrary Lie algebra (however some investigations in this direction are in progress~\cite{Ter13}).

At the same time, there always exists a special f\/inite-dimensional representation acting in the linear space of the
Lie algebra~$\mathfrak{g}$, $\tau = \ad$.
In the general case the adjoint representation $\ad$ is not faithful since the center of the Lie algebra~$\mathfrak{g}$
is a~kernel of it, $\mathfrak{z} = \ker \ad$.
Let $\{e_{\mu}\}$ be a~basis of $\mathfrak{z}$ and let the set $\{e_a\}$ forms a basis of the subspace $\mathfrak{p}$
complementary to $\mathfrak{z}$.
Fix certain canonical coordinates in the local group~$G_e$ and let these coordinates be
connected to the basis $\{e_a,e_{\mu}\}$ of the Lie algebra~$\mathfrak{g}$.
The functions $\Phi^a(x,y)$ can be found from the matrix equality
\begin{gather}
\label{e1-3-1}
\Ad_{g_x}\Ad_{g_y}=\Ad_{g_z},
\qquad
z=\Phi(x,y),
\qquad
\Ad_{e^X}=\exp(\ad_X),
\qquad
X \in \mathfrak{g},
\end{gather}
and appear to be the components of the composition function for the local quotient group $\bar{G} =
G_e/\exp(\mathfrak{z})$.
So, the problem of construction of the composition function $\Phi=(\Phi^a,\Phi^\mu)$ on~$G_e$ can be reduced to the
solution of equations~\eqref{e1-3-1} with unknown variables $z^a=\Phi^a(x,y)$ and to the computation of~$\Phi^\mu(x,y)$
for the central components.
The last problem will be solved in Section~\ref{Section4} and it will be shown that functions $\Phi^\mu(x,y)$ can be constructed by quadratures.

Finally, we discuss the important issue mentioned in the introduction: How the know\-ledge of the left- and
right-invariant vector f\/ields on a~Lie group can be used for the construction of realizations of its Lie algebra~by
vector f\/ields in f\/inite-dimensional spaces?

Consider an~$m$-dimensional space~$M$ (an open domain in $\mathbb{R}^m$) with coordinates $q = (q^1, \dots$, $q^m)$.
Let $X_i = X_i^a(q) \partial_{q^a}$ be vector f\/ields on~$M$ that realize an~$n$-dimensional Lie algebra~$\mathfrak{g}$. 
Then there exists one and only one local transformation Lie group $G_e$ of~$M$
whose Lie algebra coincides with the above realization of~$\mathfrak{g}$.
Let $U = \psi(G_e) \subset \mathbb{R}^n$ be an image of $G_e$ under coordinate mapping~$\psi$.
It means that there exists a~function $\Psi\colon M \times U \rightarrow M$ such that
\begin{gather}
\Psi(\Psi(q,x), y) = \Psi(q, \Phi(x,y)),
\qquad
\Psi(q,0) = q,
\qquad
q \in M,
\quad
x,y \in U,
\nonumber
\\
\label{e1-4-0}
X_i^a(q) = \frac{\partial \Psi^a(q,x)}{\partial x^i} \bigg|_{x = 0}.
\end{gather}
The vector f\/ields $X_i$ that are def\/ined by~\eqref{e1-4-0} are called (inf\/initesimal) \textit{generators} of the action of
the group $G_e$ on~$M$.

Suppose that the action of $G_e$ on~$M$ is \textit{transitive}.
It means that any point $q_0 \in M$ has a~neighborhood $V \subset M$ such that for any $q \in V$ there exists an element
$g_x \in G_e$ with $q = \Psi(q_0,x)$.
This implies that $\rank (X_i^a(q)) = m$, $q \in V$.
We f\/ix a~point $q_0 \in M$ and denote by $H$ the isotropy group of the point $q_0$ under the action of~$G_e$, $H = \{g_x \in G_e \,|\, \Psi(q_0,x) = q_0\}$,
which is a~subgroup of $G_e$.
As a result, we obtain a $G_e$-equivariant dif\/feomorphism between points of the space~$M$ and elements of the space
of right cosets $H {\setminus} G_e$~\cite{GorOni88}.
The choice of the point $q_0 \in M$ is not essential as far as the isotropy groups of dif\/ferent points of a homogeneous
space are conjugate.

So, the transitive action of local transformation group is def\/ined by the pair $(G_e,H)$, where $G_e$ is a~local Lie group
and~$H$ is a~subgroup of $G_e$.
This is equivalent to the assignment of the pair~$(\mathfrak{g}, \mathfrak{h})$, where~$\mathfrak{g}$ is the Lie algebra
of the group $G_e$, and $\mathfrak{h}$ is the Lie algebra of the group~$H$.
Inversely, given a Lie algebra~$\mathfrak{g}$ and its subalgebra $\mathfrak{h}$, we can construct the
corresponding local groups $G_e$ and~$H$ and the domain~$M$, where $G_e$ acts transitively; $M$ can be def\/ined as the~space
of right cosets $H{\setminus} G_e$.
Subalgebras of~$\mathfrak{g}$ that are connected by inner automorphisms correspond to equivalent actions
of the local group $G_e$, because of equivariant dif\/feomorphism of the homogeneous spaces.

Note that in general case the group action of $G_e$ on the space of right cosets $H {\setminus} G_e$ may be \textit{not
effective}, i.e., there may exist $g_x \in G_e$ such that $\Psi(q,x) = q$ for all $q \in H {\setminus} G_e$.
A~number of researchers without loss of generality restrict their consideration to the class of ef\/fective actions of
transformation groups.
For instance, if the action of $G_e$ on $H {\setminus} G_e$ is not ef\/fective, then we can consider the ef\/fective action of
the quotient group $G_e / N$ on the given homogeneous space,
where~$N$ is the maximal normal subgroup of $G_e$ that is contained in~$H$~\cite{GorOni88}.
Here we do not restrict ourselves to ef\/fective group actions and allow~$H$ to contain a~nontrivial normal subgroup
of~$G_e$.
In terms of Lie algebras, it means that the subalgebra $\mathfrak{h}$ may include nonzero ideals of the algebra
$\mathfrak{g}$.

An arbitrary element of the Lie group $G_e$ can be represented as $g_x = h_y \bar{g}_q$, where $h_y \in H$ and $\bar{g}_q$
is a~f\/ixed representative of the right coset $H g_x$,
\begin{gather*}
x = (q,y),
\qquad
x^a = q^a,
\quad
a= 1, \dots, m,
\qquad
x^{m + \beta} = y^\beta, \quad \beta = 1, \dots, n - m.
\end{gather*}
Here $q^a$ are coordinates in the space of right cosets $H {\setminus} G_e$ and $y^{\beta}$ the coordinates in the
subgroup~$H$.
Therefore, the action of the local group $G_e$ on $M \simeq H {\setminus} G_e$ is reduced to the transformation of the
coset representatives $\bar{g}_q g_z = h(q,z) \bar{g}_{\Psi(q,z)}$, where $h(q,z) \in H$ is a~\textit{factor} of the
homogeneous space.
Multiplying the last equality by $h_y$ we get
\begin{gather*}
g_{\Phi((q,y),z)} = (h_y \bar{g}_q) g_z = (h_y h(q,z)) \bar{g}_{\Psi(q,z)}.
\end{gather*}
This implies that each~$a$-th component of the composition function $\Phi(x,z)=\Phi((q,y),z)$ does not depend on the
coordinates in~$H$ and coincides with the respective component of the composition function $\Psi(q,z)$
\begin{gather}
\label{e1-4-2}
\Psi^a(q,z) = \Phi^a((q,y),z),
\qquad
a= 1, \dots, m.
\end{gather}
The equalities~\eqref{e1-4-2},~\eqref{e1-1-5} and~\eqref{e1-4-0} allow us to connect the left-invariant vector f\/ields~$\xi_i$ on~$G_e$ with the corresponding generators $X_i$ of the group action on the homogeneous space $M \simeq H {\setminus}
G_e$
\begin{gather}
\label{e1-4-1a}
\xi_i(q,y) = \xi_i^a(q) \frac{\partial}{\partial q^a} + \xi_i^{\beta}(q,y) \frac{\partial}{\partial y^{\beta}},
\\
\label{e1-4-1b}
X_i(q) = \xi_i^a(q) \frac{\partial}{\partial q^a}.
\end{gather}

Concluding this section, we would like to make the following remark.
The problems of the construction of generators of the transitive transformation group and the realization of the Lie
algebra by vector f\/ields with a given number of independent variables are connected but def\/initely are not equivalent.
The second problem is much more complicated and requires more sophisticated methods (see, for
example,~\cite{ Lyc89, Gus12, PopBoyNesLut03}).
Our paper is concentrated on the solution of the f\/irst problem.

\section[Computation of invariant vector fields and 1-forms in second canonical coordinates]{Computation of invariant vector f\/ields and 1-forms\\ in second canonical coordinates}

\looseness=1
The practical computation of components of invariant vector f\/ields and 1-forms in the f\/irst ca\-no\-ni\-cal coordinates is
a~complicated problem even for low-dimensional Lie groups.
The application of the formulas~\eqref{e1-2-2a},~\eqref{e1-2-2} to the explicit computation requires evaluation of the involved
functions at the matrices $\ad_X$ and $\exp(t\ad_{X})\exp(\ad_Y)$.
These problems are linear and they are solved by reduction of the matrices to their Jordan normal forms.
In the f\/irst canonical coordinates the matrices $\ad_X$ and $\exp(t\ad_{X})\exp(\ad_Y)$ depend on~$n$ and $2n+1$
variab\-les~$x^i$ and~$x^i$,~$y^j$,~$t$, respectively, which
makes the problem quite complicated.
If all the above calculations are done, then the result of computation is cumbersome and hardly applicable in practice.
In the second canonical coordinates the components of the invariant vector f\/ields and 1-forms are relatively simple and
can be easily calculated.
This fact is proven by the following algorithm, which originates from the work of one of the authors of the present
paper~\cite{Shi97}.

We apply the dif\/ferential of a left translation $(L_{g_x})_*$ to a~basis vector $e_k$ of the Lie algebra~$\mathfrak{g}$.
Then, taking into account the equations~\eqref{e1-1-5} that def\/ine $\xi_i^j(x)$, we get
\begin{gather*}
\left(L_{g_x}\right)_*e_k=\left(L_{g_x}\right)_*\left.\partial_{y^k} g_y\right|_{y=0}=
\left.\partial_{y^k}\left(g_xg_y\right)\right|_{y=0}=\left.\partial_{y^k}g_{\Phi(x,y)}\right|_{y=0}
\\
\phantom{\left(L_{g_x}\right)_*e_k}
=\left.\frac{\partial\Phi^i(x,y)}{\partial y^k}\right|_{y=0}\partial_{x^i}g_x=\xi^i_k(x)\partial_{x^i}g_x,
\end{gather*}
which is equivalent to the conditions
\begin{gather}
\label{e2-1}
\omega^i_k(x)e_i=\big(L_{g^{-1}_x}\big)_*\partial_{x^k}g_x.
\end{gather}

Choose the second canonical coordinates on the local group $G_e$
\begin{gather}
\label{e2-3}
g_x=g_n\big(x^n\big)\cdots g_1\big(x^1\big),
\qquad
g_i(t) \equiv \exp(t e_i).
\end{gather}
The relation $\partial_t g_k(t)|_{t=0}^{}=e_k$ obviously implies $\partial_{x^1}g_x=(L_{g_x})_*e_1$.
For any $k>1$ we obtain $\partial_{x^k}g_x=(L_{g_n})_*\cdots(L_{g_k})_*
(R_{g_1})_*\cdots(R_{g_{k-1}})_*e_k$.
In the chosen coordinate system we also have
\begin{gather*}
\big(L_{g_x^{-1}}\big)_*=\big(L_{g_1^{-1}}\big)_*\big(L_{g_2^{-1}}\big)_*\cdots \big(L_{g_n^{-1}}\big)_*.
\end{gather*}
Due to the commutativity of the right and left translations and in view of~\eqref{e1-1-6}, the conditions~\eqref{e2-1} can be rewritten~as
\begin{gather*}
\omega^i_k(x)e_i=\big[\big(L_{g_1^{-1}}\big)_*  (R_{g_1} )_*\big] \big[\big(L_{g_2^{-1}}\big)_*
 (R_{g_2} )_*\big]\cdots \big[\big(L_{g_{k-1}^{-1}}\big)_*  (R_{g_{k-1}} )_*\big]e_k
\\ \phantom{\omega^i_k(x)e_i}
{}=\Ad_{g_1^{-1}}\Ad_{g_2^{-1}}\cdots \Ad_{g_{k-1}^{-1}}e_k.
\end{gather*}
So, the components of left-invariant 1-forms in the second canonical coordinates are calculated by the formulas
\begin{gather}
\omega_1^i(x)=\delta_1^i,
\nonumber\\
\omega^i_k(x)=\big\|\exp \big({-}x^1\ad{e_1} \big)\exp \big({-}x^2\ad{e_2} \big)\cdots
\exp \big({-}x^{k-1}\ad{e_{k-1}}\big)\big\|^i_k,
\qquad
k>1.\label{e2-2}
\end{gather}

The use of~\eqref{e1-1-5b} allows us to f\/ind the components of right-invariant 1-forms and left- and
right-invariant vector f\/ields.
In view of~\eqref{e2-2}, the general structure of the left-invariant 1-forms in the chosen coordinates is
\begin{gather*}
\omega^i(x)=\delta^i_1 dx^1+\omega^i_2\big(x^1\big) dx^2+\omega^i_3\big(x^1,x^2\big) dx^3+\cdots+ \omega^i_n\big(x^1,\dots,x^{n-1}\big) dx^n.
\end{gather*}
It's obvious that $\xi_1=\partial_{x^1}$ and, if $[e_1,e_2]=0$, then also $\xi_2=\partial_{x^2}$, etc.
All the functions~$\xi_i^j$ do not depend on $x^n$ and, if the condition $[e_n,e_{n-1}]=0$ is satisf\/ied, then~$\xi_i^j$ do also not depend on~$x^{n-1}$, etc.

The suggested method of the construction of invariant f\/ields and 1-forms can be easily generalized for an arbitrary
coordinate system of the second type.
For instance, one can choose an arbitrary order of exponentials in~\eqref{e2-3}: $g_x=g_{\pi(n)}(x^{\pi(n)})\cdots
g_{\pi(1)}(x^{\pi(1)})$, where $\pi\in S_n$ is a~certain permutation of the set $\{1,\ldots,n\}$.
In this case the left-invariant f\/ield $\xi_{\pi(1)}$ is diagonal $\xi_{\pi(1)}=\partial_{x^{\pi(1)}}$.
Therefore, changing the basis of the Lie algebra we can diagonalize any given vector f\/ield along the chosen direction.

The second canonical coordinates are especially convenient for the coordinate realization of generators of the group
action on homogeneous space.
Indeed, let~$M$ be the right homogeneous space equivariant to the space of right cosets $M \simeq H {\setminus} G_e$,
let $\mathfrak{h}$ be the Lie algebra of group~$H$ with a~basis $\{e_\beta\}$ and let $\mathfrak{p}=\{e_a\}$ be a~linear
subspace complementary to the space $\mathfrak{h}$.
We choose the second canonical coordinates on~$G_e$ with
\begin{gather}
\label{e2-2a}
g_{(q,y)}=\prod \limits_{\beta = 1}^{\dim \mathfrak{h}} \exp\big(y^\beta e_\beta\big) \prod \limits_{a = 1}^{\dim
\mathfrak{p}} \exp\big(q^a e_a\big).
\end{gather}
Then the coordinate form of left-invariant vector f\/ields and generators of the transformation group is given
by~\eqref{e1-4-1a} and~\eqref{e1-4-1b}, respectively.

\looseness=1
We should emphasize that a~researcher who solves the problem of realization of a~Lie algebra by left-invariant vector
f\/ields on the associated Lie group can take into account only the commutation relations.
For instance, one can assume $\xi_1=\partial_{x^1}$ and also can choose the diagonal form for all the f\/ields commuting
with it.
After that, the system of constructed vector f\/ields is to be completed by the rest of them with unknown coef\/f\/icients; as
a~result, the overdetermined system of dif\/ferential equations is to be obtained from the commutation relations.
Consequent integration of this system gives the solution of the problem.
This procedure provides vector f\/ields, which will coincide with the left-invariant vector f\/ields in the second canonical
coordinates (up to coordinate transformations and up to a~basis change from the automorphism group).
In other words, the simplest realization of a~Lie algebra by left-invariant vector f\/ields is a~realization in the second
canonical coordinates and in that sense, this type of coordinates is privileged.

As an example, consider the six-dimensional unsolvable Lie algebra~$\mathfrak{g}$ with the following non-zero
commutation relations:
\begin{gather}
[e_1,e_2]=e_6,
\qquad
[e_1,e_4]=-e_1,
\qquad
[e_1,e_5]=e_2,
\qquad
[e_2,e_3]=e_1,
\nonumber\\
[e_2,e_4]=e_2, \qquad [e_3,e_4]=-2 e_3,
\qquad
[e_3,e_5]=e_4,
\qquad
[e_4,e_5]=- 2e_5.\label{e2-4}
\end{gather}
This algebra is a~semidirect sum of the three-dimensional nilpotent ideal and the simple Lie algebra
$\mathfrak{so}(1,2)$.

We choose the second canonical coordinates~\eqref{e2-3} $g_x = g_6(x^6)g_5(x^5)\cdots g_1(x^1)$ on the corresponding local Lie
group.
The coordinate representations of basic left-invariant 1-forms are obtained by the computation of matrix exponentials
$\exp(- x^i \ad e_i)$ according to~\eqref{e2-2},
\begin{gather}
\omega^1 = dx^1 - x^2 dx^3 + \big(x^1 - 2 x^2 x^3\big) dx^4 + x^3 e^{2 x^4} \big(x^2 x^3 - x^1\big) dx^5,
\nonumber\\
\omega^2 = dx^2 - x^2 dx^4 + e^{2 x^4}\big(x^2 x^3 - x^1\big) dx^5,
\nonumber
\\
\omega^3 = dx^3 + 2 x^3 dx^4 - \big(x^3\big)^2 e^{2 x^4} dx^5,
\nonumber
\\
\omega^4 = dx^4 - x^3 e^{2 x^4} dx^5,
\nonumber
\\
\omega^5 = e^{2 x^4} dx^5,
\nonumber
\\
\omega^6 = dx^6 - x^1 dx^2 - \frac12 \big(x^2\big)^2 dx^3 + x^2 \big(x^1 - x^2 x^3\big) dx^4 + \frac12 e^{2 x^4}\big(x^1 - x^2 x^3\big)^2 dx^5.\label{e2-4a}
\end{gather}

The matrices $\Ad_{g_x^{-1}}$ and their inverses $\Ad_{g_x}$ are constructed via the multiplication of the~mat\-rix exponentials
$\exp(-x^i \ad e_i)$ in the appropriate order, here~-- in the decreasing order of indices.
Components of the right-invariant 1-forms are given by the formula $\|\sigma(x)\|=-\Ad_{g_x}\cdot\|\omega(x)\|$ as it follows from~\eqref{e1-1-6b}.
The matrices of components of the right- and left-invariant vector f\/ields are computed as the inverses of the corresponding
matrices for 1-forms $\|\sigma(x)\|$ and $\|\omega(x)\|$, respectively.
In this example, the f\/inal expression for the left- and right-invariant vector f\/ields looks as follows (the notation $\partial_{x^i}\equiv
\partial/\partial x^i$ is assumed):
\begin{gather}
\xi_1 = \partial_{x^1},\qquad  \xi_2 = \partial_{x^2} + x^1 \partial_{x^6},
\qquad
\xi_3 = x^2 \partial_{x^1} + \partial_{x^3} + \frac12 \big(x^2\big)^2 \partial_{x^6},
\nonumber\\
\nonumber
\xi_4 = - x^1 \partial_{x^1} + x^2 \partial_{x^2} - 2 x^3 \partial_{x^3} + \partial_{x^4},
\\
\nonumber
\xi_5 = x^1 \partial_{x^2} - \big(x^3\big)^2 \partial_{x^3} + x^3 \partial_{x^4} + e^{-2 x^4} \partial_{x^5} + \frac12 \big(x^1\big)^2
\partial_{x^6},
\qquad
\xi_6 = \partial_{x^6},
\\
\nonumber
\eta_1 = -\big(e^{-x^4} + x^3 x^5 e^{x^4}\big) \partial_{x^1} - x^5 e^{x^4} \partial_{x^2} - x^2 \big(e^{-x^4} + x^3 x^5 e^{x^4}\big)
\partial_{x^6},
\\
\nonumber
\eta_2 = -e^{x^4} \big(x^3 \partial_{x^1} + \partial_{x^2} + x^2 x^3 \partial_{x^6}\big),
\qquad
\eta_3 = -e^{-2 x^4} \partial_{x^3} - x^5 \partial_{x^4} + \big(x^5\big)^2 \partial_{x^5},
\\
\eta_4 = -\partial_{x^4} + 2 x^5 \partial_{x^5},
\qquad
\eta_5 = -\partial_{x^5},
\qquad
\eta_6 = -\partial_{x^6}.\label{e2-5}
\end{gather}

Consider the four-dimensional homogeneous space $M = H {\setminus} G_e$ with the isotropy sub\-algebra
$\mathfrak{h}=\{e_4,e_5\}$, $H=\exp(\mathfrak{h})$.
Since the basis vector $e_6$ generates the center of~$\mathfrak{g}$, the ele\-ment~$g_6(x^6)$ commutes with any
element of the group,
\begin{gather*}
g_6\big(x^6\big)g_5\big(x^5\big)\cdots g_1\big(x^1\big) = g_5\big(x^5\big)\cdots g_1\big(x^1\big) g_6\big(x^6\big).
\end{gather*}
Therefore, the element $g_x$ has representation~\eqref{e2-2a} in the chosen second canonical coordinates and the gene\-ra\-tors~$X_i$
of the transformation group of homogeneous space with the local coordinates $q^1 = x^1$, $q^2=x^2$, $q^3=x^3$, $q^4=x^6$
are obtained from the left-invariant f\/ields~\eqref{e2-5} by the formal substitution $\partial_{x^1} \rightarrow
\partial_{q^1}$, $\partial_{x^2} \rightarrow \partial_{q^2}$, $\partial_{x^3} \rightarrow \partial_{q^3}$,
$\partial_{x^4} \rightarrow 0$, $\partial_{x^5} \rightarrow 0$ and $\partial_{x^6} \rightarrow \partial_{q^4}$
\begin{gather*}
X_1 = \partial_{q^1},
\qquad
X_2 = \partial_{q^2} + q^1 \partial_{q^4},
\qquad
X_3 = q^2 \partial_{q^1} + \partial_{q^3} + \frac12 \big(q^2\big)^2 \partial_{q^4},
\\
X_4 = - q^1 \partial_{q^1} + q^2 \partial_{q^2} - 2 q^3 \partial_{q^3},
\qquad
X_5 = q^1 \partial_{q^2} - \big(q^3\big)^2 \partial_{q^3} + \frac12 \big(q^1\big)^2 \partial_{q^4},
\qquad
X_6 = \partial_{q^4}.
\end{gather*}

The problem on realizations of a~Lie algebra whose commutation relations contain arbitrary parameters is quite
common in applications.
Isotropy subalgebras may also depend on arbitrary parameters.
The described method is still useful in these cases.
We illustrate this by a~simple example.

The canonical basis of the Poincar\'e algebra $\mathfrak{p}(1,3)$ is
$\{P_A, J_{AB}, A<B\}$, where $P_A$ are generators of
translations and $J_{AB}$ are generators of Lorentz transformations in the Minkowski spacetime, $A,B = 0,1,2,3$.
The complete classif\/ication of all inequivalent subalgebras of the algebra $\mathfrak{p}(1,3)$ is given in~\cite{FusBarBar91}.
Consider the four-dimensional subalgebra~$\mathfrak{g}$ from this classif\/ication with the basis elements
\begin{gather*}
e_1 = P_1,\quad
e_2 = P_2,\quad
e_3 = J_{12} + \alpha J_{03},\quad
e_4 = P_0 + P_3,
\qquad
\alpha \in \mathbb{R},
\end{gather*}
which satisfy the following nonzero commutation relations:
\begin{gather*}
[e_1,e_3]=e_2,
\qquad
[e_2,e_3]=-e_1,
\qquad
[e_3,e_4]= -\alpha e_4.
\end{gather*}
We construct a realization of the algebra~$\mathfrak{g}$
that is associated with the isotropy subalgebra $\mathfrak{h}$ spanned by the element $\{e_3+b e_4\}$.

For this purpose, we choose the second canonical coordinates $(y,q^1,q^2,q^3)$ on a
neighborhood of the identity in the Lie group with the Lie algebra~$\mathfrak{g}$ such that an arbitrary element from this neighborhood is represented as
\begin{gather*}
g=\exp(y(e_3 + b e_4)) \exp\big(q^3 e_3\big) \exp\big(q^2 e_2\big) \exp\big(q^1 e_1\big),
\qquad
b \in \mathbb{R}.
\end{gather*}
The matrix $\Ad_g$ is a~result of matrix exponentiations.
The components of left-invariant 1-forms $\omega^i_j$ are to be found on the next step and the
formulas~\eqref{e1-1-5b} and~\eqref{e1-1-6b} give expressions for left- and right-invariant vector f\/ields in the
considered coordinate system ($\partial_y \equiv \partial/\partial y$, $\partial_{q^i} \equiv \partial/\partial q^i$)
\begin{gather*}
\xi_1 = \partial_{q^1},
\qquad
\xi_2 = \partial_{q^2},
\qquad
\xi_3 = -q^2 \partial_{q^1} + q^1 \partial_{q^2} + \partial_{q^3},
\qquad
\xi_4 = (1/b) e^{-\alpha q^3}(\partial_y - \partial_{q^3}),
\\
\eta_1 = - \cos\big(y + q^3\big) \partial_{q^1} - \sin\big(y + q^3\big)\partial_{q^2},
\qquad
\eta_2 = \sin\big(y + q^3\big)\partial_{q^1} - \cos\big(y + q^3\big)\partial_{q^2},
\\
\eta_3 = \big(e^{\alpha y} - 1\big)\partial_y - e^{\alpha y}\partial_{q^3},
\qquad
\eta_4 = (1/b) e^{\alpha y}(\partial_{q^3} - \partial_y).
\end{gather*}

Finally, we can construct the generators of the transformation group whose isotropy
subgroup is associated with the subalgebra $\mathfrak{h}=\{ e_3+b e_4\}$.
As mentioned above, this can be realized by the restriction of the left-invariant vector f\/ields on the space
of functions that do not depend on the variable~$y$.
This allows us to formally substitute $\partial_y\to0$ and get
\begin{gather*}
X_1 = \partial_{q^1},\qquad  X_2 = \partial_{q^2},\qquad  X_3 = -q^2 \partial_{q^1} + q^1 \partial_{q^2} + \partial_{q^3},
\qquad  X_4 = -(1/b) e^{-\alpha q^3}\partial_{q^3}.
\end{gather*}

\section{Composition function in second canonical coordinates}\label{Section4}

We represent the Lie algebra~$\mathfrak{g}$ as the~direct sum of two subspaces~-- the center $\mathfrak{z} = \ker \ad$ of~$\mathfrak{g}$ and~a linear complement~$\mathfrak{p}$ to~$\mathfrak{z}$ in~$\mathfrak{g}$,
$\mathfrak{g} = \mathfrak{z} \oplus \mathfrak{p}$.
Let $\{e_{\mu}\}$ and $\{e_a\}$ be bases in~$\mathfrak{z}$ and in~$\mathfrak{p}$, respectively.
Since any element of $\exp(\mathfrak{z})$ commutes with any element of the local group $G_e$, in the second
canonical coordinates we get (the Einstein summation convention is not assumed)
\begin{gather}
g_x g_y = \left(\prod \limits_{\mu = 1}^{\dim \mathfrak{z}} \exp\big(x^{\mu} e_{\mu}\big) \prod \limits_{a = 1}^{\dim
\mathfrak{p}} \exp\big(x^{a} e_{a}\big) \right) \left(\prod \limits_{\nu = 1}^{\dim \mathfrak{z}} \exp\big(y^{\nu} e_{\nu}\big) \prod
\limits_{b = 1}^{\dim \mathfrak{p}} \exp\big(y^{b} e_{b}\big) \right)
\nonumber\\
\phantom{g_x g_y}
= \prod \limits_{\mu = 1}^{\dim \mathfrak{z}} \exp\big(\big(x^{\mu} + y^{\mu}\big) e_{\mu}\big) \left(\prod \limits_{a = 1}^{\dim
\mathfrak{p}} \exp\big(x^{a} e_{a}\big) \prod \limits_{b = 1}^{\dim \mathfrak{p}} \exp\big(y^{b} e_{b}\big) \right).\label{e3-1}
\end{gather}
In the general case, the subspace $\mathfrak{p}$ is not a~subalgebra of~$\mathfrak{g}$ and the expression in the last big brackets
can be rewritten as
\begin{gather}
\label{e3-2}
\prod \limits_{a = 1}^{\dim \mathfrak{p}} \exp\big(x^{a} e_{a}\big) \prod \limits_{b = 1}^{\dim \mathfrak{p}} \exp\big(y^{b} e_{b}\big)
= \prod \limits_{\mu = 1}^{\dim \mathfrak{z}} \exp\big(\Theta^{\mu}(x,y) e_{\mu}\big) \prod \limits_{a = 1}^{\dim \mathfrak{p}}
\exp\big(\bar{\Phi}^{a}(x,y) e_{a}\big).
\end{gather}
The equality~\eqref{e3-2} should be considered as the def\/inition of the functions $\Theta^{\mu}(x,y)$ and $\bar{\Phi}^{a}(x,y)$.
It is important that these functions depend only on coordinates~$x^a$ and~$y^a$  corresponding to the subspace~$\mathfrak{p}$.
The substitution of~\eqref{e3-2} into~\eqref{e3-1} gives
\begin{gather*}
g_x g_y = \prod \limits_{\mu = 1}^{\dim \mathfrak{z}} \exp \big(\big(x^{\mu} + y^{\mu} + \Theta^{\mu}(x,y)\big) e_{\mu}
\big) \prod \limits_{a = 1}^{\dim \mathfrak{p}} \exp\big(\bar{\Phi}^{a}(x,y) e_{a}\big).
\end{gather*}
Therefore, the composition function $\Phi(x,y)$ in the second canonical coordinates related to the
decomposition $\mathfrak{g} = \mathfrak{z} \oplus \mathfrak{p}$ looks as
\begin{gather}
\label{e3-4}
\Phi^{\mu}(x,y) = x^{\mu} + y^{\mu} +\Theta^{\mu}(x,y),
\qquad
\Phi^a(x,y) = \bar{\Phi}^a(x,y).
\end{gather}

It is necessary to note here that the subspace $\mathfrak{p}$ in the decomposition $\mathfrak{g} = \mathfrak{z} \oplus \mathfrak{p}$
can be chosen in many dif\/ferent ways.
This ambiguity originates from the possibility to change the basis of the Lie algebra~$\mathfrak{g}$: $e_{\mu}
\rightarrow e_{\mu}$, $e_a \rightarrow e_a + \lambda_a^{\mu} e_{\mu}$, where $(\lambda_a^{\mu})$ is an arbitrary matrix of the appropriate size.
It is easy to prove that this basis change makes the functions $\Theta^{\mu}$ and $\bar{\Phi}^a$ to
transform as
\begin{gather*}
\Theta^{\mu}(x,y) \rightarrow \Theta^{\mu}(x,y) + \big(x^{a} + y^{a} - \bar{\Phi}^a(x,y)\big) \lambda_a^{\mu},
\qquad
\bar{\Phi}^a(x,y) \rightarrow \bar{\Phi}^a(x,y).
\end{gather*}

\begin{proposition}
The composition function for a~local Lie group $G_e$ in the second canonical coordinates can be found by quadratures.
\end{proposition}

\begin{proof}
The proposition is proven in a constructive way by the demonstration of the algorithm for computing the functions
$\bar{\Phi}^a(x,y)$ and $\Theta^{\mu}(x,y)$ that are involved in the representation~\eqref{e3-4}, for the composition function
of local group $G_e$.

Since $\ad e_{\mu} = 0$, the matrix of the adjoint representation $\Ad_{g_x}$ in the second canonical coordinates
depends only on $x^a$, and hence
\begin{gather}
\label{e3-5}
\Ad_{g_x} = \prod \limits_{a = 1}^{\dim \mathfrak{p}} \exp\big(x^a \ad e_a\big).
\end{gather}
Then from~\eqref{e1-3-1} and~\eqref{e3-4} we obtain
\begin{gather}
\label{e3-6}
\prod \limits_{a = 1}^{\dim \mathfrak{p}} \exp\big(x^a \ad e_a\big) \prod \limits_{b = 1}^{\dim \mathfrak{p}} \exp\big(y^b \ad e_b\big)
= \prod \limits_{a = 1}^{\dim \mathfrak{p}} \exp\big(\bar{\Phi}^a(x,y) \ad e_a\big).
\end{gather}
Let a~tuple of $x^a$ be the coordinates in the local quotient group $G_e/\exp(\mathfrak{z})$.
Then the matrices~\eqref{e3-5} are the matrices of adjoint representation of the group $G_e$ and, at the same time, they
form a~\textit{faithful} representation of the quotient group $G_e/\exp(\mathfrak{z})$ that acts in the linear space~$\mathfrak{g}$.
Therefore, the matrix relation~\eqref{e3-6} allows us to def\/ine uniquely the functions $\bar{\Phi}^a(x,y)$, which are the
composition functions for the local quotient group~$G_e/\exp(\mathfrak{z})$.
So, the problem~\eqref{e3-4} is reduced to f\/inding the still undef\/ined functions~$\Theta^{\mu}(x,y)$.

Let $\xi_i(x) = \xi_i^j(x) \partial_{x^j}$ be the left-invariant vector f\/ields on the group $G_e$ which are written in
the second canonical coordinates, and let $\omega^i(x) = \omega^i_j(x) dx^j$ be the corresponding left-invariant
1-forms.
As shown in the previous section, in given coordinates this is a problem of linear algebra and can be solved by the
computation of matrix exponentials.
Note that the components $\xi_i^j(x)$ and $\omega^i_j(x)$ are functions only of the coordinates $x^a$ and do not depend
on the coordinates of the center $\exp(\mathfrak{z})$ (see~\eqref{e3-4} for the composition function of the local group
$G_e$).

Setting $i = a$, $k = \mu$ in~\eqref{e1-1-8} and then taking into account~\eqref{e3-4}, we obtain a~system of dif\/ferential
equations for the unknown functions $\Theta^{\mu}(x,y)$, where coordinates $x^a$ are parameters,
\begin{gather}
\label{e3-7}
\frac{\partial \Theta^{\mu}(x,y)}{\partial y^a} = \xi^{\mu}_j(\bar{\Phi}(x,y)) \omega^j_a(y).
\end{gather}
The system~\eqref{e3-7} is completely integrable since the 1-forms $\xi^{\mu}_j(\bar{\Phi}(x,y)) \omega^j_a(y) dy^a$ are
closed and the solution of this system with the initial condition $\Theta^{\mu}(x,0) = 0$ is given by the integral
\begin{gather}
\label{e3-8}
\Theta^{\mu}(x,y) = \int_0^y \xi^{\mu}_j(\bar{\Phi}(x,z)) \omega^j_a(z) dz^a.
\end{gather}
The proposition is proven.
\end{proof}

In order to illustrate results of this section, we construct an example of the composition function for the local Lie group
$G_e$ that corresponds to the six-dimensional Lie algebra~$\mathfrak{g}$ def\/ined by the commutation relations~\eqref{e2-4}.
The center~$\mathfrak{z}$ of~$\mathfrak{g}$ is one-dimensional, $\mathfrak{z} = \{e_6\}$.
Let the f\/ive-dimensional subspace $\mathfrak{p} = \{e_1, e_2, e_3, e_4, e_5 \}$ be the chosen linear complement to
$\mathfrak{z}$.

We compute the matrix exponentials $\exp(x^i \ad e_i)$ and substitute them into the matrix equali\-ty~\eqref{e3-6},
which gives a~system of algebraic equations on the unknown functions $\bar{\Phi}^a(x,y)$.
Solving this system we get
\begin{gather}
\Phi^1(x,y) = \bar{\Phi}^1(x,y) = x^1 e^{- y^4} + y^1 + y^3 e^{y^4} \big(x^2 + x^1 y^5\big),
\nonumber\\
\Phi^2(x,y) = \bar{\Phi}^2(x,y) = e^{y^4} \big(x^2 + x^1 y^5\big) + y^2,
\nonumber
\\
\Phi^3(x,y) = \bar{\Phi}^3(x,y) = \frac{e^{-2 y^4} x^3 + y^3  \big(1 + x^3 y^5\big)}{1 + x^3 y^5},
\nonumber
\\
\Phi^4(x,y) = \bar{\Phi}^4(x,y) = x^4 + y^4 + \ln \big(1 + x^3 y^5\big),
\nonumber
\\
\Phi_5(x,y) = \bar{\Phi}^5(x,y) = \frac{x^5 + e^{-2 x^4} y^5 + x^3 x^5 y^5}{1 + x^3 y^5}.\label{e3-9}
\end{gather}

We substitute the expressions for left-invariant vector f\/ields $\xi_i(x)$, dual 1-forms $\omega^i(x)$ and functions that are presented in~\eqref{e2-5},
\eqref{e2-4a} and~\eqref{e3-9}, respectively, into~\eqref{e3-8}.
The computing the resulting integral jointly with~\eqref{e3-4} gives
the sixth component of the composition function
\begin{gather*}
\Phi^6(x,y) = x^6 + y^6 + \frac{1}{2} \big(x^1\big)^2 y^5 + y^2 y^3 e^{y^4} \big(x^1 y^5 + x^2\big) + x^1 y^2 e^{- y^4} + \frac{1}{2} y^3 e^{2 y^4}
\big(x^1 y^5 + x^2\big)^2.
\end{gather*}

Concluding the example we f\/ind the components of the function $\Psi(q,z)$ that def\/ines the action of
the local group $G_e$ on the homogeneous space $M = H {\setminus} G_e$, where $H = \exp(\mathfrak{h})$ with $\mathfrak{h}=\{e_4, e_5\}$.
For this purpose, we choose $q^1=x^1$, $q^2=x^2$, $q^3=x^3$, $q^4=x^6$ as local coordinates on~$M$. Then in view of~\eqref{e1-4-2} we get
\begin{gather*}
\Psi^1(q,y) = q^1 e^{- y^4} + y^1 + y^3 e^{y^4} \big(q^2 + q^1 y^5\big),
\\
\Psi^2(q,y) = e^{y^4} \big(q^2 + q^1 y^5\big) + y^2,
\\
\Psi^3(q,y) = \frac{e^{-2 y^4} q^3 + y^3 \big(1 + q^3 y^5\big)}{1 + q^3 y^5},
\\
\Psi^4(q,y) = q^4 + y^6 + \frac{1}{2} \big(q^1\big)^2 y^5 + y^2 y^3 e^{y^4}
\big(q^1 y^5 + q^2\big) + q^1 y^2 e^{- y^4} + \frac{1}{2} y^3 e^{2 y^4} \big(q^1 y^5 + q^2\big)^2.
\end{gather*}

\section[Transition from second canonical coordinates to first canonical coordinates]{Transition from second canonical coordinates\\ to f\/irst canonical coordinates}\label{Section5}

The f\/irst canonical coordinates are universal in the sense that if one knows a~composition function, then the transition
to any type of canonical coordinates can be found and, consequently, the composition function can be represented in these coordinates.
Consider the mentioned transition to the second canonical coordinates.

Let $g_y^\mathrm{I}$ be a~group element in the f\/irst canonical coordinates $y^i$ and $g_x^{\mathrm{II}}$ is the same
element in the second canonical coordinates $x^i$.
The connection between the coordinate systems $y^i=Y^i(x)$ and $x^i=X^i(y)$, $X=Y^{-1}$, follows from the equality
\begin{gather}
\label{e4-1a}
\exp\left(\sum \limits_{i=1}^n y^i e_i\right)=\prod \limits_{i=1}^n \exp\left(x^ie_i\right).
\end{gather}
Using~\eqref{e1-2-2}, we multiply the exponentials and get
\begin{gather*}
Y^i(x)=\Phi^i(X_1,\Phi(X_2,\ldots)\ldots),
\\
X_1=\big(x^1,0,\ldots,0\big),
\quad
X_2=\big(0,x^2,0,\ldots\big),\quad \ldots, \quad X_n=\big(0,\ldots,0,x^n\big).
\end{gather*}
So, if we know the composition function $\Phi^\mathrm{I}$ in the f\/irst canonical coordinates, then the composition
function $\Phi^{\mathrm{II}}$ in the second canonical coordinates can be easily obtained,
\begin{gather*}
\Phi^{\mathrm{II}}(x,\tilde{x})=X\left(\Phi^{\mathrm{I}}(Y(x),Y(\tilde{x}))\right).
\end{gather*}
The purpose of this section is to f\/ind the connection between the f\/irst and the second canonical coordinates in the case of
unknown composition function.

Consider a~local one-parametric subgroup
\begin{gather*}
\exp\left(t Y\right)= g_{y_t}^\mathrm{I}=g_{x_t}^{\mathrm{II}},
\qquad
Y=\sum\limits_{i=1}^n y^ie_i.
\end{gather*}
The equation def\/ining this one-parametric curve in the f\/irst canonical coordinates is simple: $y^i_t=t y^i$.
Let $x_t=X(y_t)=X(ty)=\alpha(t)$ be the equation of the same curve in the second canonical coordinates.
As the zero coordinates correspond to the identity element of the group, we have
\begin{gather}
\label{e4-1}
x_t|_{t=0}^{}=\alpha(0)=0.
\end{gather}
The variables~$x$ and~$y$ are coordinates of the same group element, which 
is equivalent to the condition
\begin{gather}
\label{e4-1b}
x_t|_{t=1}^{}=\alpha(1)=x.
\end{gather}

It is well known that any one-parametric subgroup (as a~curve in a~group) is the integral trajectory of
a left-invariant (resp.\ right-invariant) vector f\/ield.
Consider the left- and right-invariant vector f\/ield with a chosen direction $Y\in\mathfrak{g}$ at the identity
element of the group, $\xi(x)=y^i\xi_i(x)$ and $\eta(x)=-y^i\eta_i(x)$.
Since the trajectory of such a vector f\/ield is uniquely def\/ined by the initial point and the associated direction at the identity
element, we can f\/ind the equation describing the one-parametric subgroup as the~solution of one of the systems of ordinary dif\/ferential equations
\begin{gather}
\label{e4-2}
\frac{d x_t}{dt} =\xi(x_t),
\\
\label{e4-3}
\frac{d x_t}{dt} =\eta(x_t),
\end{gather}
that additionally satisf\/ies the initial condition~\eqref{e4-1}.
This solution parametrically depends on
the variables $y^i$, $x_t=\alpha(t,y)=\alpha(1,ty)$.
Using~\eqref{e4-1b} we get the requested connection
$x=X(y)=\alpha(1,y)$ between the f\/irst and the second canonical coordinates.

Subtracting~\eqref{e4-3} from~\eqref{e4-2}, we obtain the integrals of motion
\begin{gather*}
y^i\big(\xi_i^k(x_t)+\eta_i^k(x_t)\big)=0.
\end{gather*}
Using~\eqref{e1-1-6b}, this formula can be written as
\begin{gather*}
\Ad_{g_{x_t}^{\mathrm{II}}}Y=Y.
\end{gather*}
The above integrals of motion simplify the integration of the system~\eqref{e4-2} or~\eqref{e4-3} but do not allow to
get all the functions $\alpha^i(t,y)$.
Show that this problem can be solved by quadratures and do not require any integration of the dif\/ferential
equations.

\begin{proposition}\label{Proposition2}
The transformation $x=X(y)$ connecting the first and second canonical coordinates are found by quadratures.
\end{proposition}

\begin{proof}
We replace the basis elements of the Lie algebra in~\eqref{e4-1a} by their adjoint representations, $e_i\to \ad e_i$,
which gives the matrix equality
\begin{gather}
\label{e4-4}
\Ad_{g^{\mathrm{I}}_y}=\Ad_{g^{\mathrm{II}}_x}.
\end{gather}
Recall that the exponentiation of a matrix is a~linear algebra problem.
Let $\mathfrak{z}$ be the kernel of the adjoint representation of the Lie algebra~$\mathfrak{g}$. Denote by $(y^\mu)$, $(x^\mu)$, $(y^a)$ and $(x^a)$
the f\/irst and the second canonical coordinates of the subgroup $\exp(\mathfrak{z})$ and
the f\/irst and the second coordinates of the local quotient group $G_e/\exp(\mathfrak{z})$, respectively.
For any element $ Z\in\mathfrak{z}$, its adjoint representation $\ad Z$ is the zero matrix. Hence the coordinates $y^\mu$ and
$x^\mu$ are not involved in~\eqref{e4-4}.
In other words, the equation~\eqref{e4-4} connects the coordinates of the local quotient group $G_e/\exp(\mathfrak{z})$,
which leads to $n-\dim\mathfrak{z}$ components of the $n$-component function $X(y)$,  $x^a=X^a(y)$.
Therefore, if the Lie algebra is centerless, then the equality~\eqref{e4-4} gives the complete solution.
Otherwise we need to construct $\dim\mathfrak{z}$ more components $X^\mu(y)$.

It is easy to show that in any system of canonical coordinates, components of  invariant vector f\/ields
and 1-forms do not depend on the coordinates~$x^\mu$ of the center.
Suppose that we have already computed the functions $X^a(y)$ using~\eqref{e4-4} and the components of left-invariant vector f\/ields in the second canonical coordinates using, e.g., techniques developed in Section~\ref{Section2}.
Then the equations of the system~\eqref{e4-2} that are associated with the center can be easily integrated,
\begin{gather*}
x_t^\mu = X^\mu(ty)=\int_{0}^t \xi^\mu(X(ty)) dt=\int_{0}^t y^i\xi^\mu_i(X(ty)) dt
\end{gather*}
since the expressions $\xi^\mu_i(X)$ do not involve~$X^\mu$.
Taking into account the condition~\eqref{e4-1b}, we get the rest of transformation components{\samepage
\begin{gather}
\label{e4-5}
x^\mu=X^\mu(y)=\int_{0}^1 y^i\xi^\mu_i(X(ty)) dt.
\end{gather}
The proposition is proven.}
\end{proof}

We illustrate Proposition~\ref{Proposition2} with the same six-dimensional Lie algebra~$\mathfrak{g}$, which is def\/ined by the commutation relations~\eqref{e2-4}.
As the center~$\mathfrak{z}$ of~$\mathfrak{g}$ is nontrivial, $\mathfrak{z}=\{e_6\}$, the matrix
equality~\eqref{e4-4} connects the f\/irst and the second canonical coordinates on the quotient
group $G_e/\exp(\mathfrak{z})$ as $x^i=X^i(y)$, $i=1,\ldots,5$. Specif\/ically,
\begin{gather*}
x^1 = y^1 \sinh (J) / J+2\big(y^2 y^3 - y^1 y^4\big)\sinh^2(J/2)/J^2,
\\
x^2 = y^2 \sinh(J)/J + 2 \big( y^2 y^3 + y^1 y^5\big) \sinh^2(J/2)/J^2,
\\
x^3 = y^3 \left(\sinh(2J)/(2J) + y^4 \sinh^2(J)/J^2 \right)/\left(\cosh(J) + y^4 \sinh(J)/J \right)^2,
\\
x^4 = \ln \left(\cosh(J) + y^4 \sinh(J)/J \right),
\\
x^5 = y^5 \sinh(J)/\left( J \cosh(J) + y^4 \sinh(J) \right),
\qquad
J := \sqrt{ \big(y^4\big)^2 + y^3 y^5}.
\end{gather*}
The expressions for invariant f\/ields~\eqref{e2-5} allow us to write down the integrand in~\eqref{e4-5} as
\begin{gather*}
y^i\xi^6_i(X(ty)) = y^2 X^1(t y) + y^3 X^2(ty)/2 + y^5 \left(X^1(ty)\right)^2/2+y^6,
\end{gather*}
where
\begin{gather*}
X^1(ty)=y^1 \sinh(t J)/J + 2\big(y^2 y^3 - y^1 y^4\big)\sinh^2(t J/2)/J^2,
\\
X^2(ty)=y^2\sinh (t J)/J + 2\big(y^2 y^3 + y^1 y^5\big)\sinh^2(t J/2)/J^2.
\end{gather*}
The integral in~\eqref{e4-5} is easily computed in this case, which gives
\begin{gather*}
x^6 = y^6 + \frac{y^1 y^2 \cosh(J) - y^1 y^2 - \big(y^2\big)^2 y^3 /2 + \big(y^1\big)^2 y^5 /2 + y^1 y^2 y^4}{J^2} +
\\
\phantom{x_6=}{}
+\frac{\sinh(2 J) \left( \big(y^2\big)^2 y^3 + \big(y^1\big)^2 y^5 \right )/4 - \sinh(J) \left ( y^1 y^2 y^4 + \big(y^1\big)^2 y^5 \right )}{J^3} +
\\
\phantom{x_6=}{}
+ \frac{\left ( -2 y^1 y^2 y^3 y^5 - \big(y^2\big)^2 y^3 y^4 + \big(y^1\big)^2 y^4 y^5 \right ) (\cosh(J) - \cosh(2 J)/4 - 3/4)}{J^4}.
\end{gather*}
We should note that the relation $x=X(y)$ is invertible and we can get the inverse mapping $y=Y(x)$ in elementary
functions as well.

\subsection*{Acknowledgements}

Authors greatly appreciate the cooperation of the editors and referees who put decent ef\/fort and amount of time to
improve the content and style of the paper.
We also want to especially thank the referees for the helpful discussions on the subject of the paper which moved our
understanding of the problem much further.
This work was supported by the Ministry of Education and Science of the Russian Federation (Project no.~3107).

\pdfbookmark[1]{References}{ref}
\LastPageEnding


\begin{thebibliography}{99}
\footnotesize \itemsep=0pt

\bibitem{AlhEngWu84}
Alhassid Y., Engel J., Wu J., Algebraic approach to the scattering matrix,
  \href{http://dx.doi.org/10.1103/PhysRevLett.53.17}{\textit{Phys. Rev. Lett.}} \textbf{53} (1984), 17--20.

\bibitem{BarShi03}
Baranovskii S.P., Shirokov I.V., Extensions of vector f\/ields on {L}ie groups
  and homogeneous spaces, \href{http://dx.doi.org/10.1023/A:1023283418983}{\textit{Theoret. and Math. Phys.}} \textbf{135}
  (2003), 510--519.

\bibitem{BasLahZhd01}
Basarab-Horwath P., Lahno V., Zhdanov R., The structure of {L}ie algebras and
  the classif\/ication problem for partial dif\/ferential equations, \href{http://dx.doi.org/10.1023/A:1012667617936}{\textit{Acta
  Appl. Math.}} \textbf{69} (2001), 43--94, \href{http://arxiv.org/abs/math-ph/0005013}{math-ph/0005013}.

\bibitem{Che46}
Chevalley C., Theory of Lie groups, Vol.~1, Princeton University Press,
  Princeton, NJ, 1946.

\bibitem{Coh65}
Cohn P.M., Lie groups, \textit{Cambridge Tracts in Mathematics and Mathematical
  Physics}, Vol.~46, Cambridge University Press, New York, 1957.

\bibitem{Dra12}
Draisma J., Transitive {L}ie algebras of vector f\/ields: an overview,
  \href{http://dx.doi.org/10.1007/s12346-011-0062-9}{\textit{Qual. Theory Dyn. Syst.}} \textbf{11} (2012), 39--60,
  \href{http://arxiv.org/abs/1107.2836}{arXiv:1107.2836}.

\bibitem{FusBarBar91}
Fushchich V.I., Barannik L.F., Barannik A.F., Subgroup analysis of Galilean and
  Poincar\'e groups and reduction of nonlinear equations, Naukova Dumka, Kiev,
  1991.

\bibitem{FusZhd97}
Fushchych W.I., Zhdanov R.Z., Symmetries of nonlinear Dirac equations,
  Mathematical Ukraina Publisher, Kyiv, 1997, \href{http://arxiv.org/abs/math-ph/0609052}{math-ph/0609052}.

\bibitem{GonShi09}
Goncharovskii M.M., Shirokov I.V., An integrable class of dif\/ferential
  equations with nonlocal nonlinearity on {L}ie groups, \href{http://dx.doi.org/10.1007/s11232-009-0149-5}{\textit{Theoret. and
  Math. Phys.}} \textbf{161} (2009), 332--345.

\bibitem{GonKamOlv92}
Gonz{\'a}lez-L{\'o}pez A., Kamran N., Olver P.J., Lie algebras of vector f\/ields
  in the real plane, \href{http://dx.doi.org/10.1112/plms/s3-64.2.339}{\textit{Proc. London Math. Soc.}} \textbf{64} (1992),
  339--368.

\bibitem{GorOni88}
Gorbatsevich V.V., Onishchik A.L., Lie groups of transformations, in Lie Groups
  and Lie Algebras, \textit{Current Problems in Mathematics. {F}undamental
  Directions}, Vol.~20, VINITI, Moscow, 1988, 103--240.

\bibitem{HauSch68}
Hausner M., Schwartz J.T., Lie groups; {L}ie algebras, Gordon and Breach
  Science Publishers, New York~-- London -- Paris, 1968.

\bibitem{HerOlv96}
Heredero R.H., Olver P.J., Classif\/ication of invariant wave equations,
  \href{http://dx.doi.org/10.1063/1.531785}{\textit{J.~Math. Phys.}} \textbf{37} (1996), 6414--6438.

\bibitem{KurShi08}
Kurnyavko O.L., Shirokov I.V., Construction of invariant wave equations of
  scalar particles on {R}iemannian manifolds with external gauge f\/ields,
  \href{http://dx.doi.org/10.1007/s11232-008-0087-7}{\textit{Theoret. and Math. Phys.}} \textbf{156} (2008), 1169--1179.

\bibitem{Lie80}
Lie S., Theorie der Transformationsgruppen, Vols.~1--3, Teubner, Leipzig, 1888,
  1890, 1893.

\bibitem{Lyc89}
Lychagin V.V., Classif\/ication of the intransitive actions of {L}ie algebras,
  \textit{Soviet Math.}  (1989), no.~5, 25--38.

\bibitem{Mil68}
Miller Jr. W., Lie theory and special functions, \textit{Mathematics in Science and
  Engineering}, Vol.~43, Academic Press, New York~-- London, 1968.

\bibitem{Mos78}
Mosolova M.V., A new formula for {$\ln (e^{A}e^{B})$} in terms of the
  commutators of the elements {$A$} and {$B$}, \href{http://dx.doi.org/10.1007/BF01431425}{\textit{Math. Notes}} \textbf{23}
  (1978), 448--452.

\bibitem{Gus12}
O'Cadiz~Gustad C., Local structure of 2 dimensional solvable lie algebra
  actions on the plane, \href{http://dx.doi.org/10.1134/S1995080212040099}{\textit{Lobachevskii~J. Math.}} \textbf{33} (2012),
  317--335.

\bibitem{Pet66}
Petrov A.Z., New methods in the general theory of relativity, Nauka, Moscow,
  1966.

\bibitem{Pon73}
Pontryagin L.S., Continuous groups, Nauka, Moscow, 1973.

\bibitem{PopBoyNesLut03}
Popovych R.O., Boyko V.M., Nesterenko M.O., Lutfullin M.W., Realizations of
  real low-dimensional {L}ie algebras, \href{http://dx.doi.org/10.1088/0305-4470/36/26/309}{\textit{J.~Phys.~A: Math. Gen.}}
  \textbf{36} (2003), 7337--7360, \href{http://arxiv.org/abs/math-ph/0301029}{math-ph/0301029}.

\bibitem{ShaShi95}
Shapovalov A.V., Shirokov I.V., Noncommutative integration of linear
  dif\/ferential equations, \href{http://dx.doi.org/10.1007/BF02065973}{\textit{Theoret. and Math. Phys.}} \textbf{104}
  (1995), 921--934.

\bibitem{Shc06}
Shchepochkina I.M., How to realize a {L}ie algebra by vector f\/ields,
  \href{http://dx.doi.org/10.1007/s11232-006-0078-5}{\textit{Theoret. and Math. Phys.}} \textbf{147} (2006), 450--469,
  \href{http://arxiv.org/abs/math.RT/0509472}{math.RT/0509472}.

\bibitem{Shi97}
Shirokov I.V., Construction of {L}ie algebras of f\/irst-order dif\/ferential
  operators, \href{http://dx.doi.org/10.1007/BF02766382}{\textit{Russian Phys.~J.}} \textbf{40} (1997), 525--530.

\bibitem{Ter13}
Terzis P.A., Faithful representations of Lie algebras and homogeneous spaces,
  \href{http://arxiv.org/abs/1304.7894}{arXiv:1304.7894}.

\end{thebibliography}
\end{document}